\documentclass[conference]{IEEEtran}
\IEEEoverridecommandlockouts
\usepackage{cite}
\usepackage{amsmath,amssymb,amsfonts}
\usepackage{algorithmic}
\usepackage{graphicx}
\usepackage{textcomp}
\usepackage{multirow}
\usepackage{caption}
\usepackage{url}
\usepackage{booktabs, siunitx}
\usepackage{multicol}
\usepackage[bookmarks=false]{hyperref}
\usepackage{float}
\usepackage{xcolor}
\def\BibTeX{{\rm B\kern-.05em{\sc i\kern-.025em b}\kern-.08em
        T\kern-.1667em\lower.7ex\hbox{E}\kern-.125emX}}
\begin{document}
\begin{NoHyper}
    
    \title{Blockchain-Based Transferable Digital Rights of Land

    }
    
    \author{\IEEEauthorblockN{1\textsuperscript{st} Ras Dwivedi}
        \IEEEauthorblockA{\textit{Computer Science and Engineering} \\
            \textit{Indian Institute of Technology, Kanpur}\\
        }
        \and
        \IEEEauthorblockN{2\textsuperscript{nd} Sumit Patel}
        \IEEEauthorblockA{\textit{Computer Science and Engineering} \\
            \textit{Indian Institute of Technology Kanpur}\\
        }
        \and
        \IEEEauthorblockN{3\textsuperscript{rd} Prof. Sandeep Shukla}
        \IEEEauthorblockA{\textit{Computer Science and Engineering} \\
            \textit{Indian Institute of Technology Kanpur}\\
        }
    }
    
    \maketitle
    
    \begin{abstract}
    Land, being a scarce and valuable resource, is in high demand, especially in densely populated areas of older cities. Development authorities require land for infrastructure projects and other amenities, while landowners hold onto their land for both its usage and its financial value. Transferable Development Rights (TDRs) serve as a mechanism to separate the development rights associated with the land from the physical land itself. Development authorities acquire the land by offering compensation in the form of TDRs, which hold monetary value.

In this paper, we present the tokenization of development rights, focusing on the implementation in collaboration with a development authority. While there have been previous implementations of land tokenization, we believe our approach is the first to tokenize development rights specifically. Our implementation addresses practical challenges related to record-keeping, ground verification of land, and the unique identification of stakeholders. We ensure the accurate evaluation of development rights by incorporating publicly available circle rates, which consider the ground development of the land and its surrounding areas.

    \end{abstract}
    

\section{Introduction}

Land, being a valuable asset, is typically held for the long term and is highly sought after for its development potential. Rapid urbanization in densely populated areas has led to a shortage of land, particularly in older parts of cities. The scarcity of land in these central areas poses a significant challenge for development authorities in creating new parks, widening roads, and implementing other infrastructure projects. However, due to their prime location, these lands carry substantial value, and owners are often reluctant to part with even the smallest portions. While it is possible to acquire land through legal means, the high compensation costs associated with it, combined with financial constraints faced by development authorities, make it impractical to acquire large amounts of land.

Transferable Development Rights (TDRs) present a mechanism that attempts to address both aspects of this problem. By decoupling the physical land from its development potential, TDRs allow the conversion of development potential into a digital tradable asset, akin to corporate shares. Unlike shares, TDR holders possess rights solely for development purposes and do not have ownership rights over the physical land.

The conversion of development potential into transferable development rights also offers a solution to the financial challenges faced by development authorities. Compensation for land can be provided in terms of TDRs instead of monetary value. This allows development authorities to acquire land without burdening their balance sheets.

This paper proposes the tokenization of TDRs on the blockchain. We argue that through tokenization and fair, transparent trading on the blockchain, users can obtain the fair market value of the land. Moreover, we contend that digitized development rights are well-suited for tokenization as they represent fully digital assets, backed by the guarantees of the development authority. To support our claims, we examine various cases of asset tokenization, where physical assets were tokenized using non-fungible tokens (NFTs). However, we note that the linkages between the physical asset and the digital token are often weak and pose significant challenges in the tokenization process. In our implementation, TDRs represent rights backed by development authorities and are not directly linked to any physical asset until their utilization. We model the entire life cycle of TDRs, from generation and transfer to their utilization, on the blockchain. Our ultimate goal is to create a marketplace for TDRs on the blockchain, facilitating efficient and transparent land development finance.
This paper is organized as follows: First we present the background on the blockchain and the asset tokenization highlighting why the linkages between the digital token and the physical asset is the problem. Then we present the TDR and its lifecycle. We then present the architecture of the out implementation. 

\section{Background}
Land is a limited resource that is constantly in demand. Urban planning bodies require land for new infrastructure projects and amenities such as road widening and parks. At the same time, tenants want to hold onto land for its usage and investment potential, especially in densely populated areas of cities.

While land can be easily tokenized as an asset, its ownership comes with separate possession rights that need to be enforced. This becomes particularly critical in countries where titular ownership is not established, and the state does not guarantee land possession. Simply digitizing land creates problems because it is challenging to establish a connection between the physical land and its digital representation.

When creating NFTs for physical assets like land, it's important to consider that the token may not always accurately depict the ground reality, especially in cases where the land and its surrounding area undergo development or significant changes. The value and characteristics of land can be influenced by various factors such as urbanization, infrastructure projects, zoning regulations, or environmental considerations. These dynamic aspects can lead to shifts in the physical landscape, affecting the actual state and potential of the land. Therefore, it is crucial to recognize that the tokenized representation of land may not fully reflect the current or future conditions of the property.

Transferable Development Rights (TDRs) are a mechanism used by urban development bodies to separate land from its development potential and acquire the land in exchange for these rights. These development rights are supported by the urban development bodies and are not tied to specific parcels of land. Since the value of land is closely linked to its development potential, these development rights become highly valuable digital assets.

In this context, we propose the tokenization of TDRs as an asset. TDRs are not associated with any specific land, eliminating the problem of backward linkage. This paper presents a practical implementation of the tokenization of development rights for the city of Kanpur, India, offering a solution to the challenges posed by land ownership and digitization.

\section{System Architecture}

\subsection{Overview of Quorum blockchain}

\begin{figure*}
    \centering
    \includegraphics[width=0.9\textwidth]{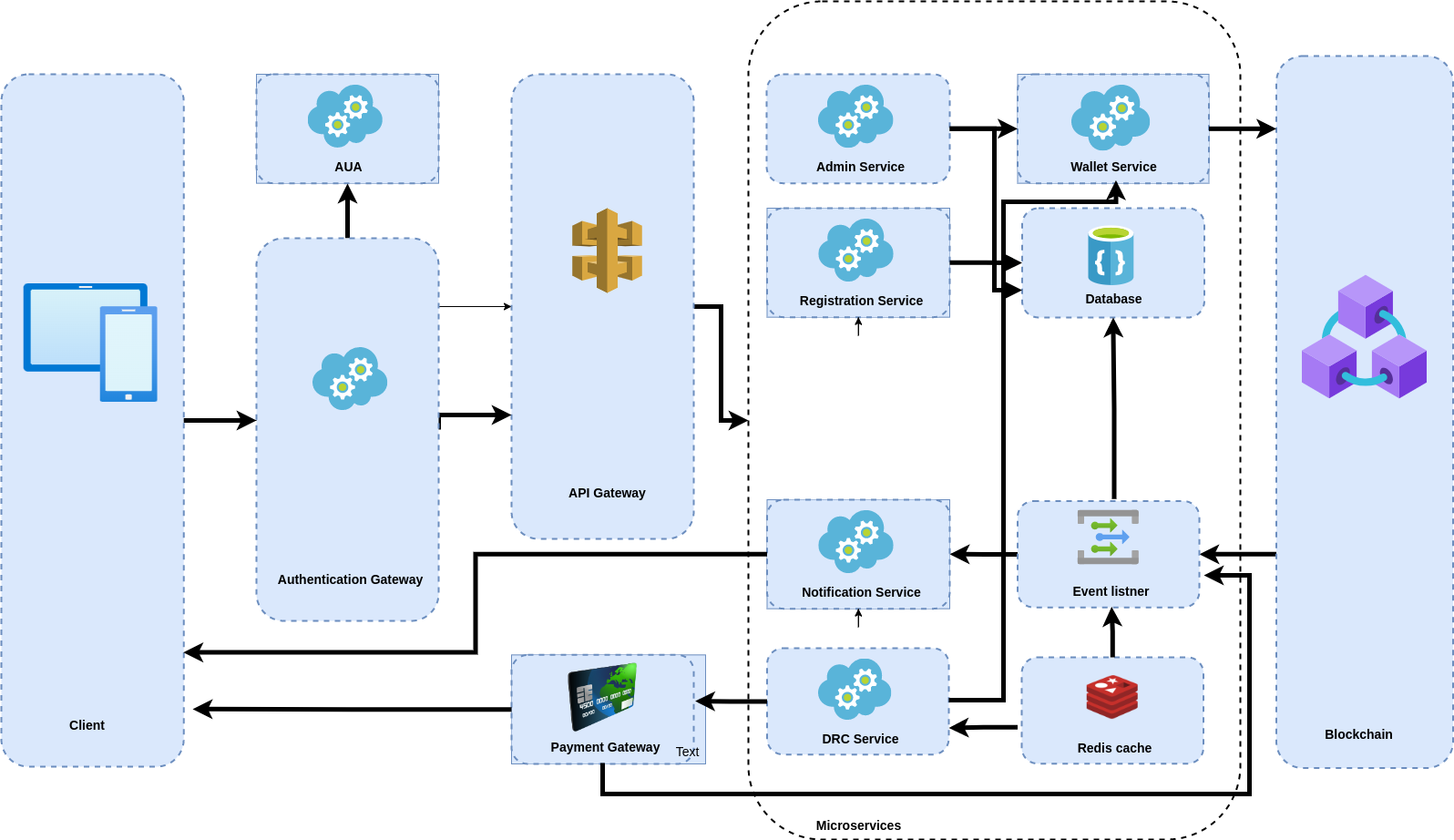}
    \caption{Technical Architecture}
    \label{fig:system architecture}
\end{figure*}

{\it} Blockchain is a type of distributed ledger technology (DLT) designed to facilitate secure, transparent, and immutable transactions. Quorum, a variant of this technology, is a private, permissioned blockchain based on Ethereum. It enables private transactions among specific users, managing network and peer permissions to ensure that only approved nodes participate in the Quorum blockchain network.

{\it} Quorum utilizes a consensus algorithm based on voting, and it offers an Istanbul Byzantine Fault Tolerance (IBFT) consensus mechanism. As it's compatible with the Ethereum Virtual Machine (EVM), Quorum also supports smart contracts programmed in Solidity, making it a versatile and secure option for managing complex transactions and contracts within a permissioned network.

\subsection{NFT standard for DRCs}
{\it} We have used the ERC721 protocol for issuing DRCs. ERC721 protocol, a prevalent standard on the Ethereum blockchain, is chiefly employed for non-fungible tokens (NFTs). NFTs, which are distinctive and non-divisible, are used to represent various unique digital assets such as collectibles, digital artwork, virtual real estate, and gaming items. The characteristic feature of ERC721 tokens is their distinctiveness and limited availability; each token holds a specific value and is non-interchangeable on a like-for-like basis, unlike fungible tokens.The protocol establishes a collection of functions and events that facilitate the creation, ownership, and transferability of non-fungible assets. 

{\it} Our Transferable Development Rights (TDR) application, based on blockchain, is developed using the Quorum blockchain and consists of four nodes. These nodes are hosted on separate Linux servers strategically located at different sites to avoid simultaneous downtime

{\it} In our current permissioned blockchain network configuration, we've disabled "Tessera," which is Quorum blockchain's default private transaction manager. We took this step because our application does not process any private transactions on the blockchain. We've also set the block mining time to five minutes, meaning that any new block that emerges will be auto-mined by the nodes.

The major components of our application include:

    1) Blockchain network.
    
    2) Quorum Transaction Signer.
    
    3) HashiCorp Vault.
    
    4) Backend Infrastructure.
    
    5) Smart contracts.
    
    6) User Identification via Aadhaar

\subsection{User Identification and onboarding}
During the user onboarding process, we ensure the authenticity and uniqueness of each user through Aadhaar verification. This verification helps us eliminate duplicate users from our system. During the signup phase, users provide their details along with their Aadhaar number. User verification is conducted via eKYC using Aadhaar, which involves sending a one-time password (OTP) to the Aadhaar-linked phone number. The user's details are then compared to the information obtained from Aadhaar. To maintain the privacy and security of Aadhaar, the Aadhaar number is stored in the HashiCorp Vault using a 16-digit reference ID. This reference ID is stored in MongoDB, associated with the user's ID, to establish the connection between the user and their Aadhaar.
Using Aadhaar also facilitates the process of resetting a user's password in case they forget it. During password reset, an OTP is sent to their registered mobile number for authentication. The user must input this OTP to reset their password, ensuring an added layer of security for account recovery. \\
For user sign-up, we ask him to upload his aadhaar number, if aadhaar number is found to be invalid, the sign-up attempt is automatically denied at time of eKYC, also with unique aadhaar number , no 2 users can have same aadhaar number on our platform.This features removes the malicious or unwanted users from registering to our platform. Once the user provides all the necessary registration details, the form is forwarded to an administrator for approval. Upon successful verification by the administrator, the user receives a confirmation email signifying their successful onboarding to the platform.

\subsection{Quorum Transaction Signer}
The Quorum Transaction Signer is an independent service that retrieves a user's private key from the HashiCorp Vault and signs transactions on their behalf.We have implemented a secure storage system for private keys in the transaction signer. Each user's private key is encrypted and stored in the HashiCorp Vault. To add an extra layer of protection, the keys are encrypted using the user's password. This means that even if the database is compromised, the user's keys remain safe and inaccessible to unauthorized parties.When a user needs to fetch their private key, they are prompted to enter their password. At the same time, the back-end server sends the corresponding reference ID associated with the user. Using this reference ID, the encrypted private key string is retrieved from the vault. In real time, the private key is decrypted using the user's password and made available for use.\\
One of the primary reasons for creating an independent transaction signer is the inherent risk of users losing their keys and passwords. If a private key is lost, establishing ownership of the associated asset becomes impossible. To mitigate this risk, we implemented a system that links a user's assets to their user ID, which, in turn, is connected to their user account through a smart contract. By establishing these linkages, we ensure that ownership of the asset can be maintained even if the user loses their private key. Furthermore, the user's account in the transaction signer is linked to their Aadhaar, a unique identification number. In the event that a user forgets their password, they can regain ownership of their assets through a validation process using Aadhaar. This validation updates the user's account in the smart contract, allowing them to regain control over their assets.

\subsection{Bifurcation of On-chain and Off-chain Data}
We bifurcate the storage of data between on-chain and off-chain was made to optimize the efficiency and security of our system. User specific details such as user photos, documents, contact information, and address, are stored off-chain. This approach helps reduce the burden on the blockchain network and improves the performance of the system.

On the other hand, all the crucial and relevant information concerning the land and its transactions is stored on-chain. By storing TDR (Transferable Development Right)application details, notice details, FAR (Floor Area Ratio) and DRC (Digital Rights Certificate) details on-chain, we ensure the transparency, immutability, and integrity of the data. This allows for easy verification and auditing of land-related transactions, enhancing trust among participants.

Moreover, storing user data off-chain helps protect the privacy of individuals, as sensitive personal information is not exposed on the public blockchain. By securely storing user-related details separately, we maintain compliance with data protection regulations and reduce the risk of unauthorized access or misuse.











\section{Implementation}


\begin{figure}
    \centering
    \includegraphics[width=1\linewidth]{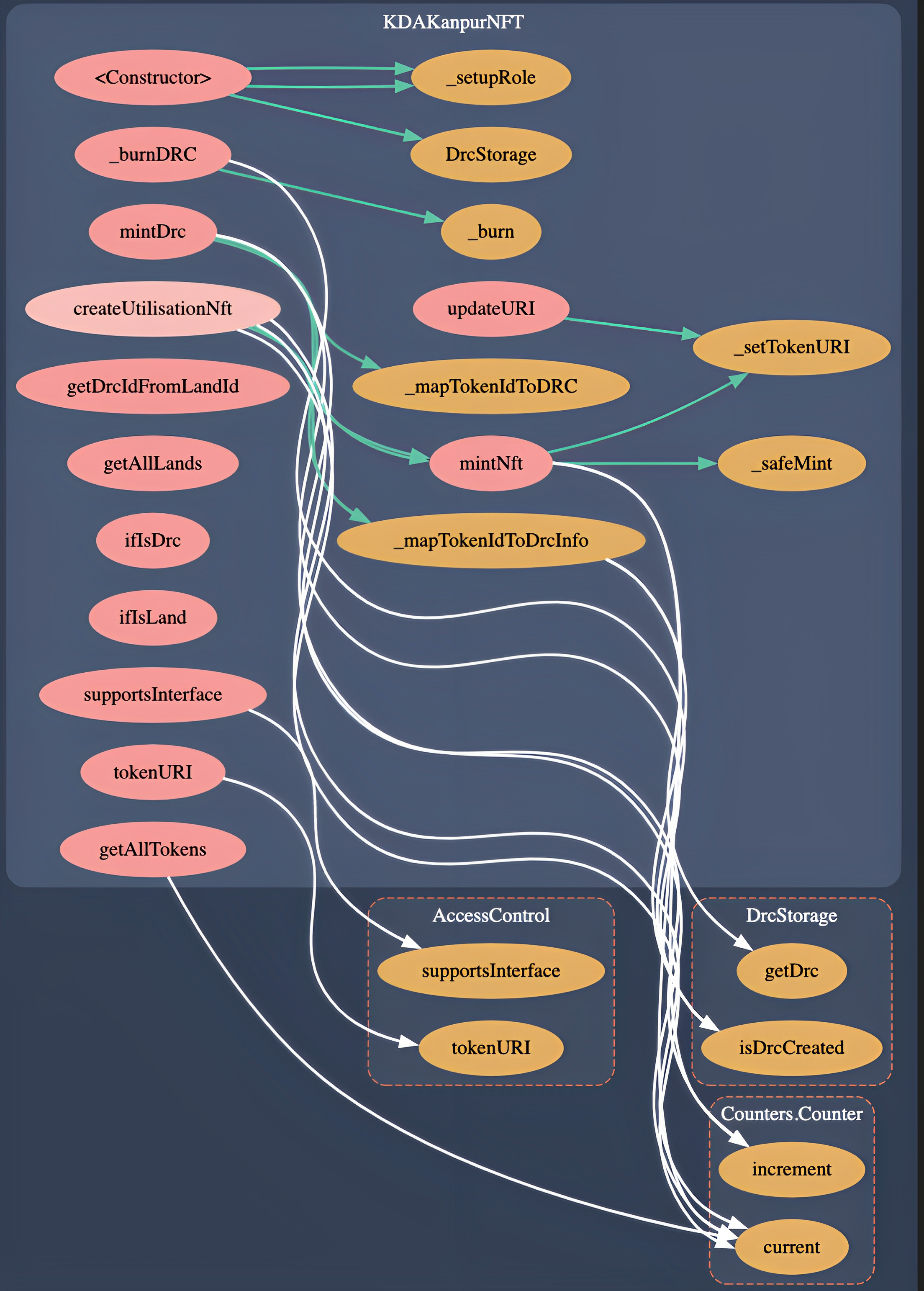}
    \caption{NFT contract strucuture}
    \label{fig:enter-label}
\end{figure}


Registered user can submit an application for the issuance of the Transferable Development Rights (TDR) if they have been notified by the development authority about the acquisition of their land. All application for the TDRs must cite the notice id against which they are applying. Upon filing a TDR application, the user is assigned a unique application number, which is stored alongside its corresponding state in our TDR smart contract. This smart contract tracks and maintains the state of each TDR ID. The TDR application undergoes a verification process by various sub-departments within the development authority. If any discrepancies are found, such as mismatches between the submitted land ownership details and the authority's system, the application is rejected or sent back for correction. 

On successful verification, the TDR application status is updated as "VERIFIED" by development authority officials from there user portal and corresponding state is updated on the blockchain through the smart contract. The TDR smart contract enables the tracking of the application's progress, ensuring accountability for any unauthorized actions by officials as signatures of officials involved in the approval/rejection on the TDR application can be traced, holding individuals responsible if necessary.Every step of the verification process requires input from development authority officials, and they must sign the transaction  on the blockchain. Smart contract maintains three status for the application: REJECTED if the application is rejected, SENT\_BACK\_FOR\_CORRECTION is the application requires certain correction, or additional documents to substantiate the claim, and VERIFIED if the application is verified by the authority. 
When an application receives approval from the officials, the user becomes qualified to receive their Development Rights Certificate (DRC). The status of this DRC ID is subsequently updated in the storage contract. Our NFT smart contract cross-verifies the present state of that DRC ID from the storage layer of the smart contract. If the DRC ID is identified in an approved state, an NFT for that DRC can be issued. However, only a user with an admin role can initiate this issuance process.

NFT of DRC contains comprehensive information pertaining to the specific land it represents. Each land detail, structured in the form of a schema, is stored on the IPFS network.The unique URL or hash generated by IPFS serves as the URI, which is subsequently stored within the NFT and transferred to the user. Consequently, each NFT possesses a distinct set of land details corresponding to the specific piece of land owned by the user. \\

\begin{table}[h]
\centering
\begin{tabular}{|c|c|c|}
\hline
\textbf{Parameter} & \textbf{Type} & \textbf{Description} \\
\hline
drcId & bytes32 & Identifier of the DRC \\
\hline
farAvailable & uint256 & FAR (Floor Area Ratio) available for allocation \\
\hline
landCount & uint256 & Total count of sub-divided lands \\
\hline
owner & address & Owner of NFT \\
\hline
lands & mapping & Mapping of land sub-divisions \\
\hline
\end{tabular}
\caption{Details stored in NFT as URI}
\label{tab:my_label}
\end{table}

Furthermore, each NFT is signed by a land development authority officer's blockchain account, thereby establishing and validating the authentic ownership of the land. This signature serves as irrefutable proof of ownership, allowing the NFT holder to demonstrate their rightful ownership of the land to any interested party.The aforementioned details are retrieved from the DRC contract and subsequently transmitted to the NFT contract. 

\subsection{Issuance and management of TDR as NFT}

{\it} The contract inherits from the ERC721, ERC721 URIStorage, and AccessControl contracts from the OpenZeppelin library. The ERC721 contract provides the basic implementation of the ERC721 standard for non-fungible tokens. The ERC721URIStorage contract extends ERC721 by including a URI for each token, which points to a JSON file with more information about the land. We have also implemented  access control logic to contracts which provides a access layer, allowing certain operations to be performed only by addresses with specific roles. For issuance of NFT contract we have used Counters library from Open-Zeppelin to keep track of token IDs. Each new token that is minted is assigned a unique ID by incrementing a counter. This counter manages that no two NFT can have the same Id. This brings the uniqueness to all the NFTs that is being issued.

{\it} When an owner develops a parcel of land for which they hold development rights, the corresponding DRC for that NFT is considered utilized. At this point, the NFT token associated with the DRC is 'burned' via the burnDRC function in our smart contract. This action can only be executed by an account assigned with the DEFAULT ADMIN ROLE, typically held by officials from the development authority. This burning process is typically initiated after physical inspection of the land to verify its developed state. Upon burning the NFT, it is removed from both the mapDRCIdToTokenId and mapTokenIdToDRCId mappings, marking the end of the lifecycle for that particular NFT and its associated development rights.

\subsection{Transfer and ownership tracking of TDR as Non Fungible Token}

The ownership and transfer of Transferable Development Rights (TDRs) are tracked via tokens using the ERC721 standard for non-fungible tokens (NFTs) on the Quorum blockchain. Here's how it works:

When a TDR is created, a corresponding NFT is minted. The NFT is unique and represents the TDR. The ownership of the NFT (and therefore the TDR) is assigned to the address that created the TDR. This process is typically handled by a function in the smart contract, as safeMint in the provided NFT contract.  The owner of an NFT is the address that currently has control over it. This is tracked on the blockchain, and the ERC721 standard provides a function ownerOf that allows anyone to query the owner of a specific NFT (and therefore the owner of the corresponding TDR).This can be used to prove ownership of development rights for that particular piece of land.

{\it} So whenever a user want to transfer his rights of land that he owns to different person, who can be potential buyer in form of real estate developer or any land authority, he can transfer his NFT, which will require 2 inputs which are public address of new buyer and the current token-id of NFT.This is handled by "transferFrom" function inside our NFT smart contract. Once the NFT for the that DRC is transferred to new owner, the current owner of that DRC can be checked with "ownerOf" function in NFT smart contract. This operation necessitates the use of a token ID as an input, which in turn provides the current owner's address for that specific NFT. Therefore, the new owner can affirm that they hold the development rights to a particular land parcel. This introduces transparency into the process by ensuring that no two owners can claim the same land parcel and preventing any user from illicitly claiming the development rights to a land parcel they do not actually own. Initially in our implementation of application,  DRC can only be claimed for certain sending zones, and can only be utilized inside certain receiving zones. 

{\it} In order to track each NFTs , we have a mapping inside our smart contract that associates each DRC ID with a token ID, and a mapping mapTokenIdToDRCId that associates each token ID with a DRC ID. These mappings allow the contract to keep track of which DRCs are associated with which tokens. The mappings within the smart contract offer an immutable record of TDR ownership, associating each NFT with a corresponding DRC. This enhances transparency, facilitates transactions, and enables easy verification and traceability of ownership history, revolutionizing the management and trade of TDRs.

\section{Conclusion}

In conclusion, issuing Development Rights Certificates (DRCs) as Non-Fungible Tokens (NFTs) on the blockchain presents a compelling solution to the challenges faced in land registration processes. By leveraging blockchain technology, transparency and trust are enhanced, ensuring the integrity of land ownership records.

The implementation of smart contracts streamlines the TDR application process, automating tasks, and facilitating efficient verification and approval procedures. The immutability of the blockchain ensures a transparent and auditable record of the TDR application's lifecycle, tracking its progress through various stages and holding officials accountable for their actions.The inclusion of unique identifiers, such as URIs, within the NFTs guarantees the accurate representation of land details and provides indisputable proof of ownership.

This transformative approach reduces disputes, promotes trust among stakeholders, and establishes a reliable system for tracking and verifying land ownership, ultimately improving the overall integrity of the land registration process.

\section{Future Work}

We are currently expanding the scope of our application by developing an on-chain marketplace specifically designed for users to trade their DRCs in a permissioned environment. This marketplace aims to provide a seamless and user-friendly platform for DRC transactions.
Furthermore, we are committed to optimizing our smart contracts and conducting bench-marking exercises to enhance their performance. 

By improving the efficiency and throughput of our smart contracts, we aim to increase the Transactions Per Second (TPS) for our application. This optimization process will contribute to a more robust and scalable system, enabling smoother and faster transaction processing.

\vspace{2mm}

\end{NoHyper}

\end{document}